\documentclass[]{spie}  

\usepackage{amsmath,amsfonts,amssymb}
\usepackage{graphicx}
\usepackage{siunitx}
\usepackage[colorlinks=true, allcolors=blue]{hyperref}
\DeclareSIUnit\angstrom{\text{\AA}}

\title{Calibration at elevation of the WEAVE fibre positioner}

\author[a]{Sarah Hughes}
\author[a,b]{Gavin Dalton}
\author[c]{Kevin Dee}
\author[c]{Don Carlos Abrams}
\author[b]{Kevin Middleton}
\author[a]{Ian Lewis}
\author[b]{David Terrett}
\author[d]{Alfonso L. Aguerri}
\author[c,d,e]{Marc Balcells}
\author[b]{Georgia Bishop}
\author[f]{Piercarlo Bonifacio}
\author[g]{Esperanza Carrasco}
\author[h]{Scott C. Trager}
\author[i]{Antonella Vallenari}

\affil[a]{Dept. of Physics, Keble Road, University of Oxford, OX1 3RH, UK}
\affil[b]{RAL Space, Science and Technology Facilities Council, Rutherford Appleton Laboratory, Harwell Oxford, OX11 OQX, UK}
\affil[c]{Isaac Newton Group, 38700 Santa Cruz de La Palma, Spain}
\affil[d]{Instituto de Astrof\'isica de Canarias, 38200 La Laguna, Tenerife, Spain}
\affil[e]{Dep. de Astrof\'isica, Universidad de La Laguna, 38200 La Laguna, Spain}
\affil[f]{GEPI, Observatoire de Paris, Université PSL, CNRS, Place Jules Janssen, 92195 Meudon, France}
\affil[g]{Instituto Nacional de Astrof\'isica, Optica y Electronica (INAOE), Mexico}
\affil[h]{Kapteyn Instituut, Rijksuniversiteit Groningen, Postbus 800, 9700 AV Groningen, Netherland}
\affil[i]{Osservatorio Astronomico di Padova, INAF, Vicolo Osservatorio 5, 35122, Padova, Italy}

\authorinfo{Further author information: (Send correspondence to Sarah Hughes)\\E-mail: sarah.hughes@physics.ox.ac.uk}

\begin{document} 
\maketitle

\begin{abstract}
WEAVE is the new wide-field spectroscopy facility\cite{Marc}$^{,}$\cite{dalton2012}$^{,}$\cite{Dalton2020} for the prime focus of the William Herschel Telescope in La Palma, Spain. Its fibre positioner is essential for the accurate placement of the spectrograph’s 960 fibre multiplex.
We provide an overview of the recent maintenance, flexure modifications, and calibration measurements conducted at the observatory prior to the final top-end assembly. This work ensures that we have a complete understanding of the positioner's behaviour as it changes orientation during observations. All fibre systems have been inspected and repaired, and the tumbler structure contains new clamps to stiffen both the internal beam and the retractor support disk onto which the field plates attach. We present the updated metrology procedures\cite{Hughes} and results that will be verified on-sky.
\end{abstract}

\keywords{fibre spectroscopy, WEAVE, WHT, robotic positioners, high multiplex spectroscopy, metrology, calibration}

\section{INTRODUCTION}
\label{intro}  

\begin{figure}[ht!]
    \centering
    \includegraphics[width=9cm]{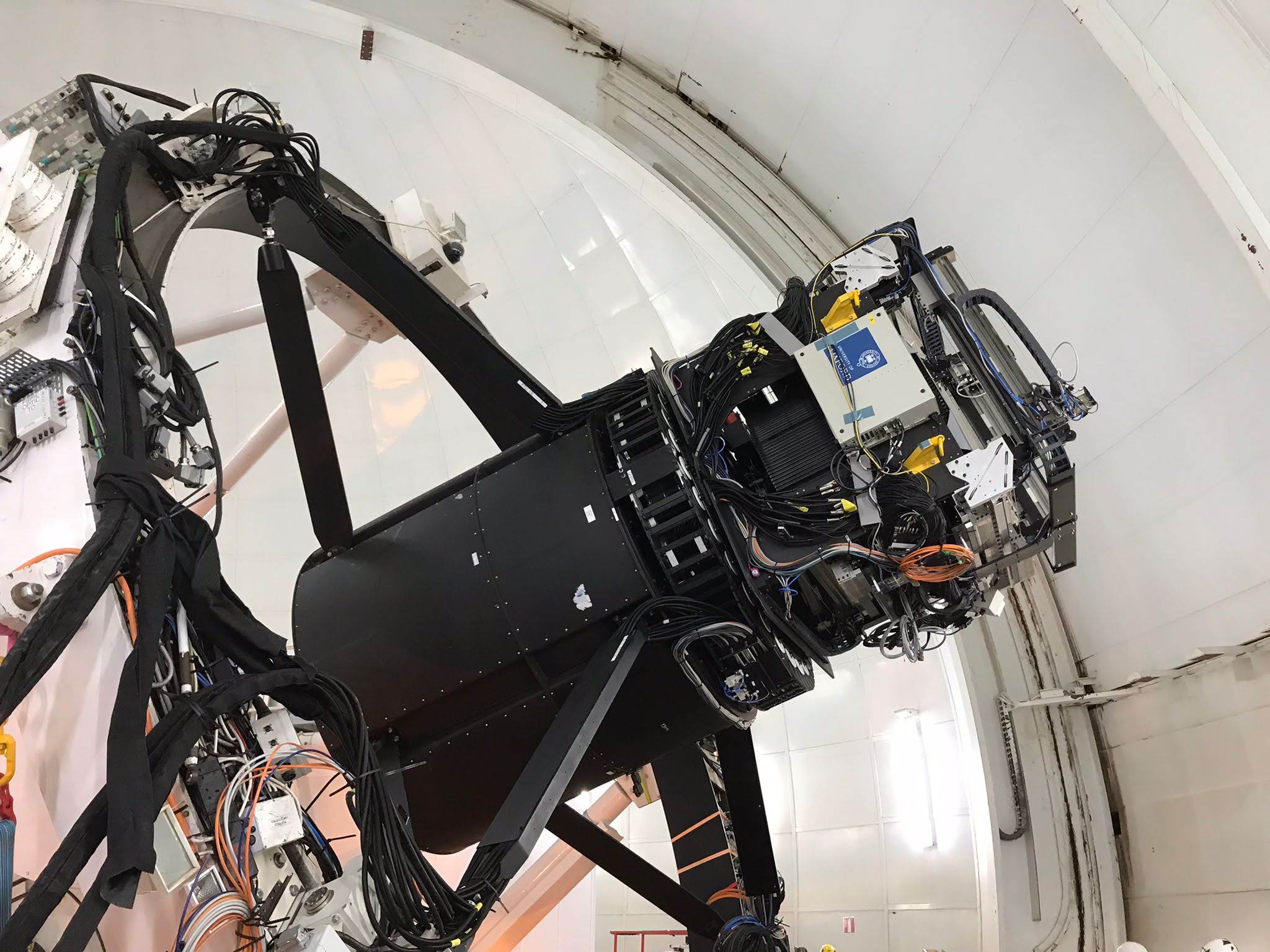}
    \caption{The assembled top end of WEAVE on the William Herschel Telescope in La Palma, Spain. This photo was taken on the $25^{\rm{th}}$ of May 2022.}
    \label{top_end}
\end{figure}

\begin{figure}[ht!]
    \centering
    \includegraphics[width=0.8\linewidth]{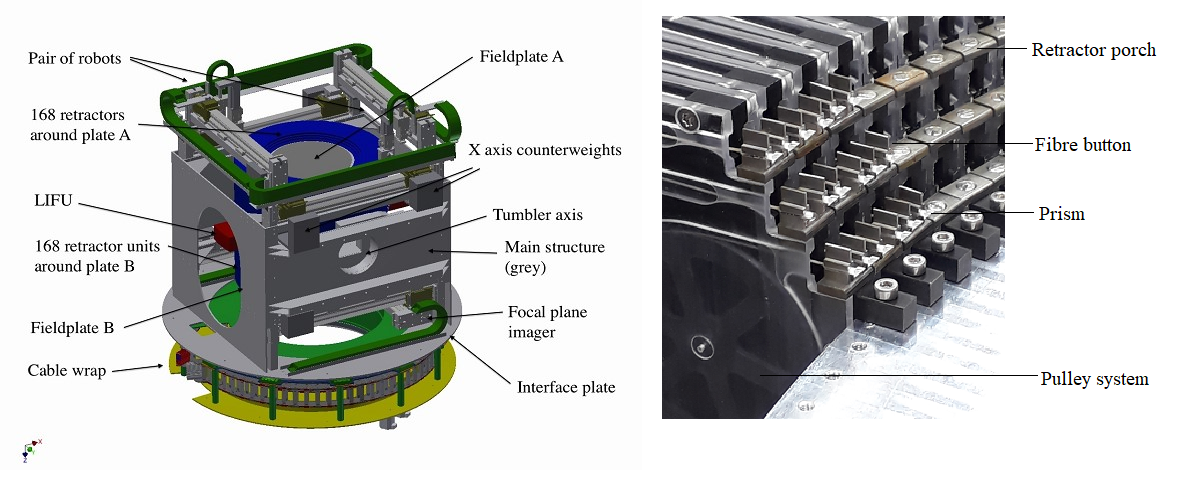}
    \caption{$\left(\rm{\emph{Left}} \right)$ A schematic diagram of the positioner, which identifies its main components. Some design modifications have been made, specifically the counterweight system, however this is outside the scope of this paper. $\left(\rm{\emph{Right}} \right)$ An image of several retractors, highlighting their tiered structure and the park positions of the MOS fibres.}
    \label{pos_diag}
\end{figure}

WEAVE is the new spectroscopy facility\cite{Dalton16} for the William Herschel Telescope (WHT) on La Palma. WEAVE has a multiplex of 960 optical fibres and a 2° field of view. It consists of four distinct subsystems: the prime focus system\cite{top-end}, the spectrograph system\cite{spectrograph}, the optical fibres\cite{fibre_tests}, and the fibre positioner\cite{Dalton2020}$^{,}$\cite{Hughes}.
The prime focus system consists of a six lens wide-field corrector with atmospheric dispersion compensator\cite{ADC} (ADC), an instrument rotator\cite{WRS} (WRS), and a new top end structure incorporating focus/tilt adjustment to the spider vanes. The ADC components comprise two air-spaced prismatic doublets which counter-rotate to correct for dispersion along the direction of the parallactic angle. The corrector prescription delivers $< 0.6"$ polychromatic images over a flat focal surface. The positioner is mounted to the field rotator, and the corrector is fixed to the top end structure.
WEAVE has three fibre observing modes: the multi-object spectrograph (MOS) fibres, the mini integral field units (mIFUs), and the large integral field unit (LIFU). Each of the 20 mIFUs is a bundle of 37 fibres, whilst the LIFU has 547 central fibres and 8 bundles around the edge with 7 fibres for sky subtraction. The positioner has two 410 mm diameter field plates (plate A and plate B) located on either side of the central tumbler, that can rotate 180° between the two. Each field plate has an independent set of fibres, which allows the next field to be configured by the robots whilst the other field plate is being used for observations. This minimises the overhead time between observations. 
The positioner uses pick-and-place technology to arrange the optical fibres in the focal plane according to the targets allocated in the field of view. Two robots (called Nona and Morta) move along x, y, z, and $\theta$-stages to pick up individual fibres arranged around the edge of the field plate in retractors. These are shown in figure \ref{pos_diag}. A gripper system attached to the each robot is used to place the fibres in their desired position, to within ±8 µm. Attached to the side of each gripper unit is an optical system that allows the imaging of the fibres (gripper camera) and field plate (upper focus). A detailed description of the positioner's design and operational capabilities can be found in Dalton et al. (2012, 2016, \& 2020)\cite{dalton2012}$^{,}$\cite{Dalton16}$^{,}$\cite{Dalton2020}, and the top-end assembly is shown in figure \ref{top_end}. A complete summary of the WEAVE survey and the early testing of the data pipeline processing systems can be found in Jin et al. (2022, \emph{in preparation})\cite{shoko} .

The dual-arm spectrograph system\cite{spectrograph} has continuous wavelength coverage (3660-9590 $\si{\angstrom}$) in the low-resolution mode (R $\sim 5750 \si{\angstrom}$) and coverage over several narrow regions (either 4040-4650 or 4730-5450, and 5950-6850 $\si{\angstrom}$) in the high-resolution mode (R $\sim 21000 \si{\angstrom}$). The spectrograph is not located on the top-end of the WHT, but on the Nasymth platform. This has significant advantages; such as improving the stability of the system when making an observation, and reducing restrictions on the size of the spectrograph due to weight balance. 
This paper will focus on the positioner and its fibre sub-systems, specifically the maintenance and calibration work completed in preparation for on-sky operations.

\section{Positioner maintenance}

After the positioner's arrival at the observatory in January 2021, we quickly realised that it had been damaged during it's journey. We suspect that it experienced a large shock, due to several (small) bolts, attaching some elements to the main structure, had been sheared off. This resulted in many months being spent detaching and reassembling the central tumbler structure, as well as a complete redesign of the robot counterweight system. The details of this procedure is outside the scope of this paper, and will instead focus on the flexure behaviour that was observed after the tumbler reassembly. We then describe the process of extracting, inspecting, and repairing all 336 retractors in the positioner, as many were disturbed as a result of the transport and maintenance work.

Once the positioner's main structure had been repaired and reassembled, it was attached to the WEAVE rotator system (WRS) via an interface plate. Together they were then placed on a testing rig that allowed us to shift its elevation angle manually. The result is a system that can mimic the positioner's orientation on the telescope.
\newpage
\subsection{Flexure behaviour}
\label{flex_tolerances}
The focus adjustment of the WHT was designed to meet a number of requirements such as the thermal expansion and flexure of the top-end structure. However, the maximum tilt and position of the field plates is driven by the allowed defocus across each fibres button when the field plate is in the observing position. The tolerances to meet this defocus budget are:
\begin{itemize}
    \item The tilt of the field plates must be less than $\pm0.0045^{\circ}$.
    \item The separation between the field plate in the observing position and the WRS interface plate should be $273.2\pm0.025$ mm.
\end{itemize}
To remain within the focus tolerances, it is crucial that we understand how the structure of the positioner will move when its orientation is changed. In the design phase, a Finite Element Analysis (FEA) model was created from the positioner's 3D Computer Aided Design (CAD) model. The FEA model was used to calculate the flexure due to varying the gravitational loading on the telescope. The tumbler itself was modelled as a series of solid bodies; the central beam, the retractor support disk (RSD) drum, the field plates, the motor, and the bearing. The model considered the combined flexure of these elements.

\begin{figure}[ht!]
    \centering
    \includegraphics[width=0.8\linewidth]{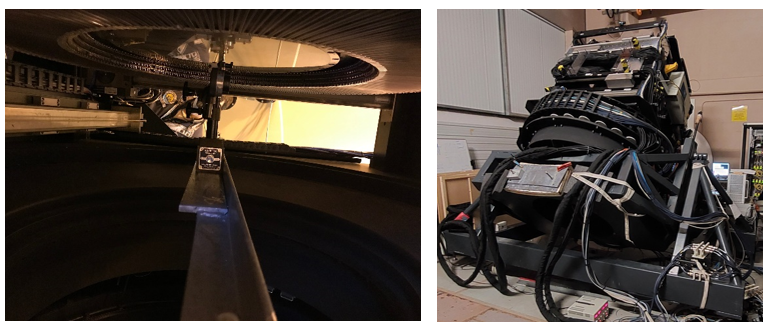}
    \caption{$\left( \rm{\emph{Left}}\right)$ The arrangement for a single DTI to measure the shift in position of the observing field plate with respect to the WRS interface plate, as the positioner's orientation is changed. It is attached to an angle iron spanning the width of the interface. The tilt was measured with two DTIs and the distance to the interface was measured with a depth micrometer, where it is zeroed at ZD $0^{\circ}$. $\left(\rm{\emph{Right}}\right)$ The positioner mounted to the WRS, on a testing rig that allows the system to change elevation.}
    \label{DTI_pic}
\end{figure}


The flexure of the positioner was measured once it was mounted to the WRS and placed on a test rig designed for the WRS. The testing rig can adjust its elevation angle up to $90^{\circ}$ from its zenith position, referred to as ZD $90^{\circ}$. Previously, ZD $0^{\circ}$ had been used for all metrology measurements in the Oxford laboratory\cite{Hughes} .

\subsubsection{Initial measurements}

A laser tracker was used to measure the positioner's metrology with respect to the WRS/POS interface plate, which was used as a fixed reference point. Four magnetic Spherically Mounted Retroreflector (SMR) nests were attached to the field plate in the observing position. The placement of the nests form a cross along the $x$ and $y$ axes of the field plates. 


\begin{table}[ht!]
\centering
\begin{tabular}{l|l|l|}
\cline{2-3}
                                                          & Plate A         & Plate B         \\ \hline
\multicolumn{1}{|l|}{$x$ direction tilt}                  & $0.009^{\circ}$ & $0.046^{\circ}$ \\ \hline
\multicolumn{1}{|l|}{$y$ direction tilt}                  & $0.024^{\circ}$ & $0.135^{\circ}$ \\ \hline
\multicolumn{1}{|l|}{Movement of plate's centre position} & 0.12 mm         & 2.293 mm        \\ \hline
\end{tabular}
\caption{The results of the tilt and shift in position of each field plate when in the observing orientation, relative to the WRS interface.}
\label{plate_tilts}
\end{table}

The measurements of the tilt and position of each field plate relative to the WRS interface with the laser tracker at ZD $0^{\circ}$ is summarised in table \ref{plate_tilts}. When comparing them to the requirements listed in section \ref{flex_tolerances}, it is clear that neither of the field plates are in the correct position.



\begin{figure}[ht!]
    \centering
    \includegraphics[width=12cm]{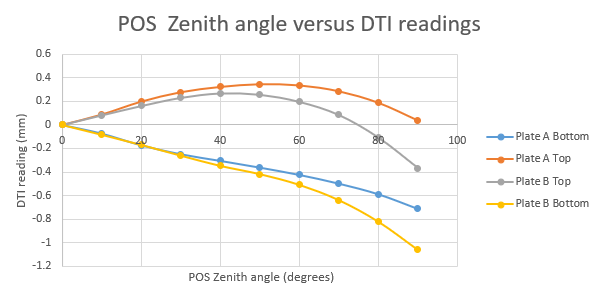}
    \caption{The displacement as a function of elevation, measured using two DTI's at both the top and bottom positions on the field plate. The bottom measurements show a continual increase in the displacement, with a maximum value of $\sim1000$ $\mu$m at ZD $90^{\circ}$. A curved profile can be observed for the top position, where the displacement increases to a maximum value of $\sim400$ $\mu$m before moving back towards its original value.}
    \label{two_DTIs}
\end{figure}

To examine the flexure behaviour further, two digital Dial Test Indicator (DTI) gauges were attached to an angle iron that spans the width of the testing rig and lies close to the field plates in the observing position. An image of this setup can be seen in figure \ref{DTI_pic}. They were each positioned at opposing sides of the field plate. For plate A, the maximum displacement was 200 $\mu$m at ZD $50^{\circ}$, with a curved profile. For plate B, the maximum displacement was $\sim$1000 $\mu$m at ZD $90^{\circ}$ with a gradual linear trend. The results are shown in figure \ref{two_DTIs}.





The magnitude of the flexure meant that mechanical interventions were needed to reduce this movement. Further testing revealed the central tumbler beam and drum as the most likely candidates for the observed displacement.

\begin{figure}[ht!]
    \centering
    \includegraphics[width=0.8\linewidth]{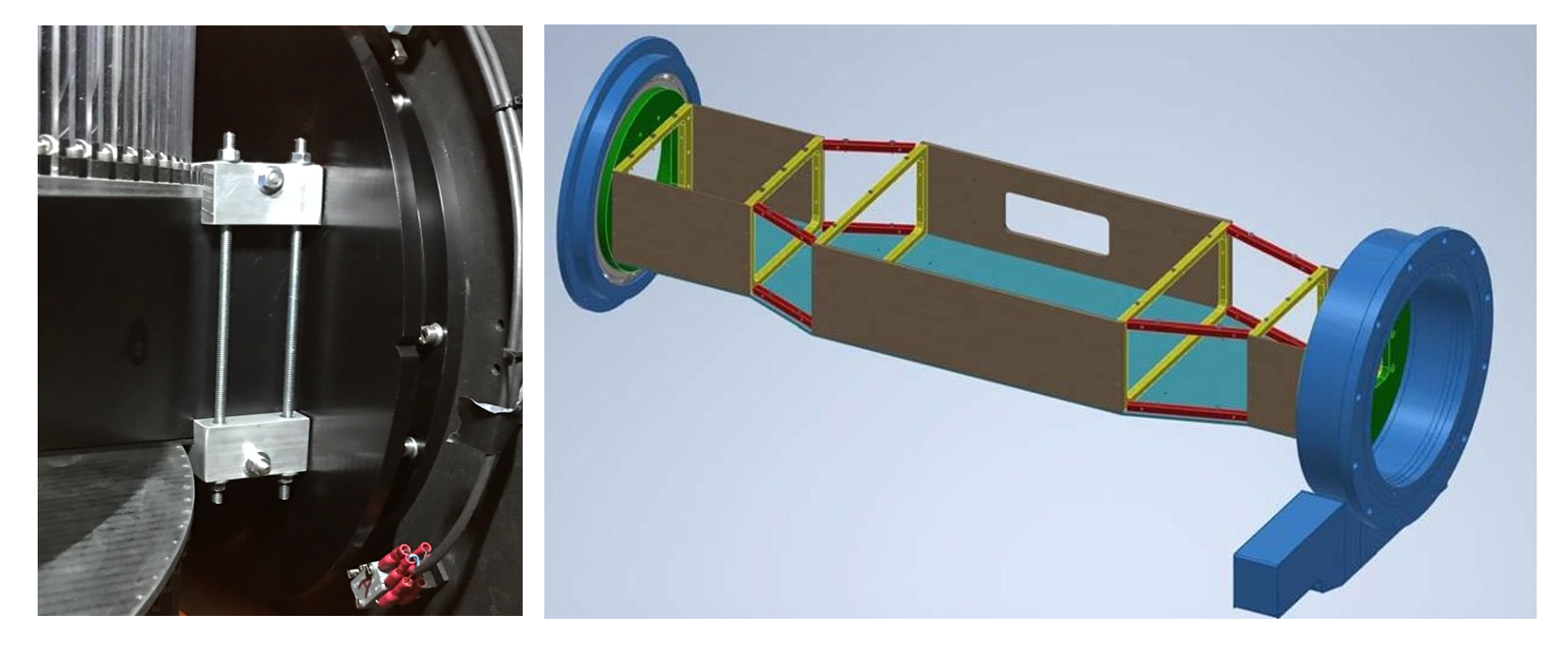}
    \caption{$\left(\rm{\emph{left}} \right)$ The clamp inserted around the bearing side of the tumbler beam, to prevent it from twisting as the positioner's orientation was changed. $\left(\rm{\emph{right}} \right)$ A schematic diagram showing the reinforcement bars, which hold in place an aluminium flat along the top of the tumbler beam. These additional structures increase the beams rigidity, which has been weakened by the bolt holes made in the side panels.}
    \label{clamp_and_bars}
\end{figure}

\subsubsection{Prevention actions}

We believe that the additional bolt holes made in the beam side plates during the tumbler's reassembly are responsible for reducing its overall stiffness. An aluminium flat, attached with reinforcement brackets, were inserted inside the tumbler beam to increase its rigidity. A clamp was designed and fitted to the inside the tumbler structure, around the edge of the beam on this side to restrict its movement. These elements resulted in a reduction in the displacement magnitude from $\sim$1000 $\mu$m to 400 $\mu$m for plate B in the observing position.



\begin{figure}[ht!]
    \centering
    \includegraphics[width=12cm]{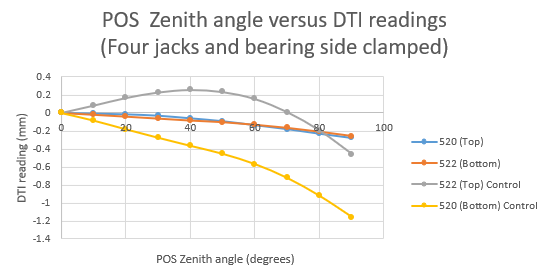}
    \caption{Shows the displacement as a function of elevation, when both the clamp, stiffening additions, and four additional jacks are used to hold the tumblers structure. Here the term \emph{Control} refers to the measurements taken prior to any insertions to reduce the flexure. We see that the additional components significantly reduces the magnitude of the displacement, and removes the curved profiles observed for each plate.}
    \label{flexure_final_results}
\end{figure}

The clamp and reinforcement brackets reduced the curvature observed to the inside of plate A's profile however, there was still a significant difference between the measurements from each DTI. Four jacks were inserted between the top plates of the tumbler drum, with the results shown in figure \ref{flexure_final_results}. The jacks were replaced by a reinforcement strut that attaches to the top plate and RSD on each side of the tumbler. This is to reduce the mass budget of the positioner. 

\subsubsection{Summary of flexure measurements}

Initial measurements made using both a laser tracker and DTI gauges showed that neither plate A or B were positioned correctly with respect to the WRS interface, and far exceeded the allowed tolerances for the telescope focusing capabilities. A clamp, an aluminium flat with its corresponding reinforcement bracket, and two RSD struts were installed into the tumbler's structure to reduce the magnitude of the displacement. The final results are shown in figure \ref{flexure_final_results}, however there is still a maximum displacement of $200$ $\mu$m, which could potentially restrict the range of positions that can be used during observations. This will be analysed during on-sky commissioning.


\subsection{Final focus position}

Despite the preventative measures in place, we still observe some flexure behaviour that restricts the range of orientations that can be used when observing, to remain within the focus tolerance. Calculations have been made as to how the final flexure results will affect the point spread function (PSF) observed with WEAVE. Previous estimations do not account for any variation in co-planarity of the field plates or any tilt, meaning that they are underestimating the residual sum of squares (RSS) of defocus. Here the updated RSS, $|\Delta x_{\rm{Tot}}|$, is defined as
\begin{equation}
    |\Delta x_{\rm{Tot}}| = \sqrt{\Delta x^{2}_{FP} + \Delta x^{2}_{0}},
\end{equation}
where $\Delta x^{2}_{FP}$ is the measured field defocus and $\Delta x^{2}_{0}$ is the RSS of defocus without the affect of the field plates.
\begin{figure}[ht!]
    \centering
    \includegraphics[width=10cm]{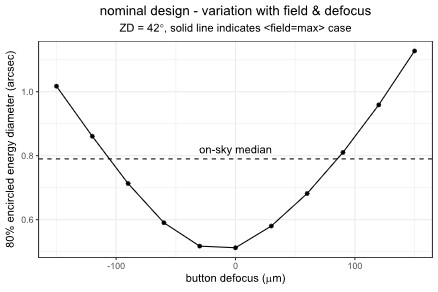}
    \caption{The encircled energy diameter of the PSF as a function of button defocus, having accounted for the plate position and tilt in the new focus budget. We can see that the worst case scenario with nominal seeing conditions is $1.2"$ at $+150$ $\mu$m defocus.}
    \label{defocus}
\end{figure}

The previous calculation assumes a symmetrical 2D gaussian profile for the PSF. This assumption underestimates the energy in the wings of the profile, and does not allow the full-width half-maximum (FWHM) to be well defined. As a result, a Moffat function has been used for these calculations, which is defined as
\begin{equation}
    f(x, y, \alpha, \beta) =  2\left(\frac{\beta - 1}{\pi \alpha^{2}}\right)\left[ 1 + \frac{ x^{2}+ y^{2}}{\alpha^2} \right]^{-\beta}.
\end{equation}
where $\beta$ and $\alpha$ are seeing dependent parameters. 

The result of modelling the PSF as a Moffat function with a FWHM of $0.8"$ and a $\beta$ value of 3, then convolving it with typical seeing conditions at the observatory is shown in figure \ref{defocus}. The worst case scenario is a FWHM of $1.2"$ at $+150$ $\mu$m defocus. The allowed angular travel range of the WRS system when observing is $\pm155^{\circ}$ rotation, and these results do not imply that any orientations need to be avoided. However, the effect of the flexure behaviour will need careful consideration, particularly as we receive results from commissioning and first light.

\subsection{Retractor interventions}

After the positioner's reassembly, remote testing began in the summer of 2021. A handful of fibres had already been disabled in the Oxford laboratory, however this list began to increase rapidly during operational testing at the observatory. The reason for these failure modes was unclear, as there was no indication that they were linked to any systematic failures or that they were concentrated to a specific region of the plate. Included in the new list of disabled fibres were a small number of guide fibres. These needed to be brought back online, as each plate only has 8 available and at least 2 are required for successful telescope guiding. A detailed description of the retractors design and assembly process can be found in Schallig E. (2019)\cite{schallig_2019}.

\begin{figure}[ht!]
    \centering
    \includegraphics[width=8cm]{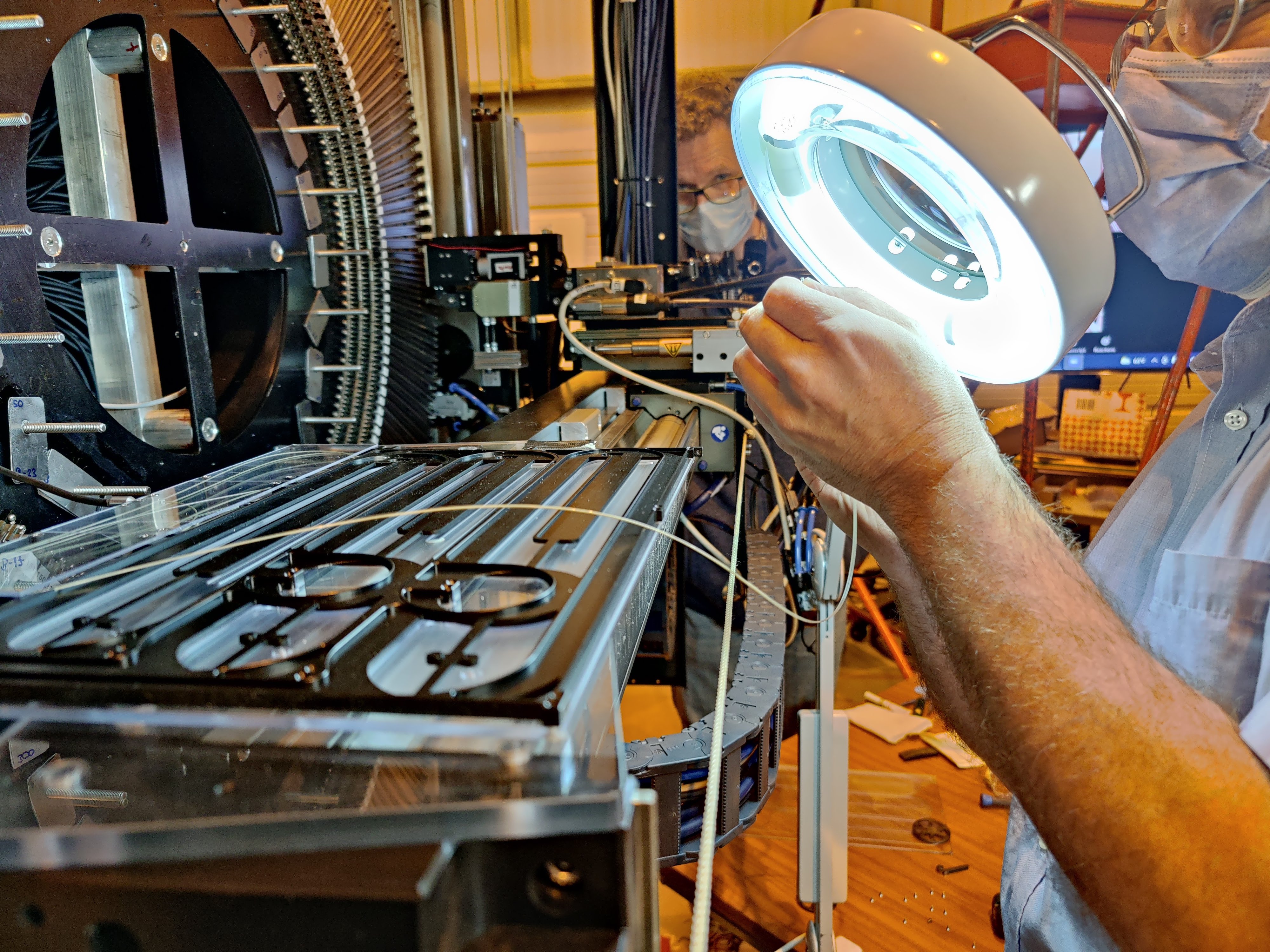}
    \caption{An image of the retractor repair process at AP3, of a guide fibre retractor on plate A. Due to the limited distance the guide retractor could be moved, the robot's Y-gantry was used to safely support the retractor during the repairs.}
    \label{retractor}
\end{figure}

In January 2022, the retractor inspection and repairs process began at the observatory after the flexure work was completed. An important aspect of this process was to understand if the retractors would be easily accessible when on the telescope. To check this, the orientation of the positioner on the rig was changed to a rotation angle of $0^{\circ}$ and an elevation angle of $90^{\circ}$ from zenith. This position corresponds to a standard telescope maintenance position called Access Park 3 (AP3). 

At AP3, we confirmed it is possible to extract both the field plates and the retractors, provided that the covers can be removed appropriately, whilst also leaving the shims under the field plate in position. The most effective way to remove the retractors is in blocks of $\sim 20$, for additional maneuvering space. It is necessary to do this when extracting retractors close to either the motor or the bearing.
The majority of the retractors have enough length in their fibre cables to allow the retractors to be laid on a nearby work surface. The exception is the guide fibre retractors, which have a reduced freedom of movement. These retractors had to be repaired using a robot gantry as the repair surface. This can be seen in figure \ref{retractor}.

After inspecting several retractors on plate A, it became clear that the entire subsystem would need to be examined. While many of the retractors had not failed, a large proportion of the pulley systems were not performing to the standards expected. This is partially due to gradual improvements in the quality control during assembly, as we increased our knowledge of how the pulleys should behave. However, it is not clearly understood how the shock affected the retractors. From assembly, we know that the retractors are quite sensitive to how their axles are positioned in the covers and aluminium plate. It is possible that the shock disrupted many of the retractors this way, or changed the operation of those which were deemed functional, yet were on the border of acceptable performance.

In total, it took approximately 5 weeks to fully complete the inspection and repairs of all 336 retractors. Many of which were not a direct result of retractor failures. As each fibre is fed into a metal button, it is surrounded by a thick tubing called a ferrule, which runs through the length of the button and ensures that the prism rests a set distance away from the front of the button. Several fibres were found to have loose ferrules which had to be re-attached carefully by-hand, as well as some magnets coming loose from their buttons. It is important to note that recovering fibres with loose ferrules can be completed at AP3 without extracting any retractors.

Since completing these repairs, there have been a small number fibres which have failed during the calibration process. This is not entirely unexpected, and we have still succeeded in reducing the total number of disabled fibres, and preventing the potential failure modes of many more. When isolating the recently disabled fibres to the retractor repairs alone, we do not observe any patterns with previous list of disabled fibres. 


\section{Metrology measurements}

Due to the modifications made to the positioner, the metrology measurements and calibration procedures completed in Oxford and described in Hughes et al.\cite{Hughes} had to be repeated. This includes re-measuring the base z-height of the field plates, evaluating the park heights of every tier 3 fibre, and new tasks such as measuring the maps and park heights as a function of position.

Many of the methods used previously have been altered to improve their efficiency, whilst adapting to the reduced positioner access as it changes orientation. 

\subsection{Field plate maps}

\begin{figure}
    \centering
    \includegraphics[width=9cm]{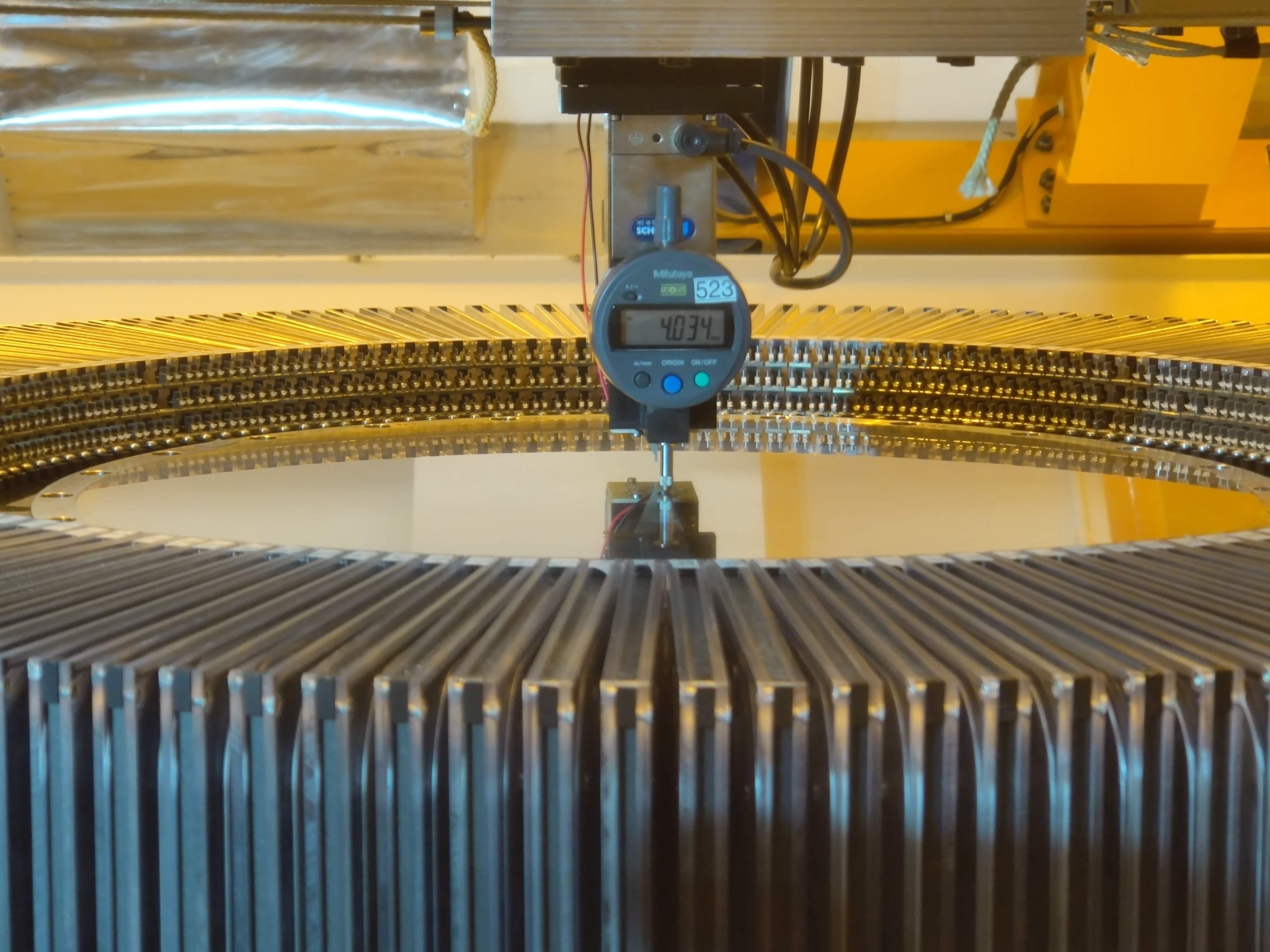}
    \caption{The DTI gauge used to measure the z-height of the field plates, for a range of orientations of the positioner. The gauge has been mounted in place of the theta counterweight system, meaning that every measurement must be made at a fixed orientation. The gauge is zeroed at coordinates (0, 0, 37) mm on the field plate.}
    \label{DTI}
\end{figure}

The field plate maps were measured using a dial gauge mounted to each robot, the exact set up can be seen in figure \ref{DTI}. The position of the gauge is in place of the new theta counterweight system, therefore it is offset from the grippers $x$ position by $\pm 55$ mm depending on the robots orientation. This means that the movement of the robots across the field plate must be restricted to prevent a collision between the robot gripper unit and the retractors.

Maps must be measured for each plate using both robots. The robots are driven to the same set of coordinate points to maintain consistency. To measure a complete map, the robot must take the majority of the measurements at $-90^{\circ}$ rotation, and the remaining points at $+90^{\circ}$ rotation.

\begin{figure}[ht!]
    \centering
    \includegraphics[width=0.9\linewidth]{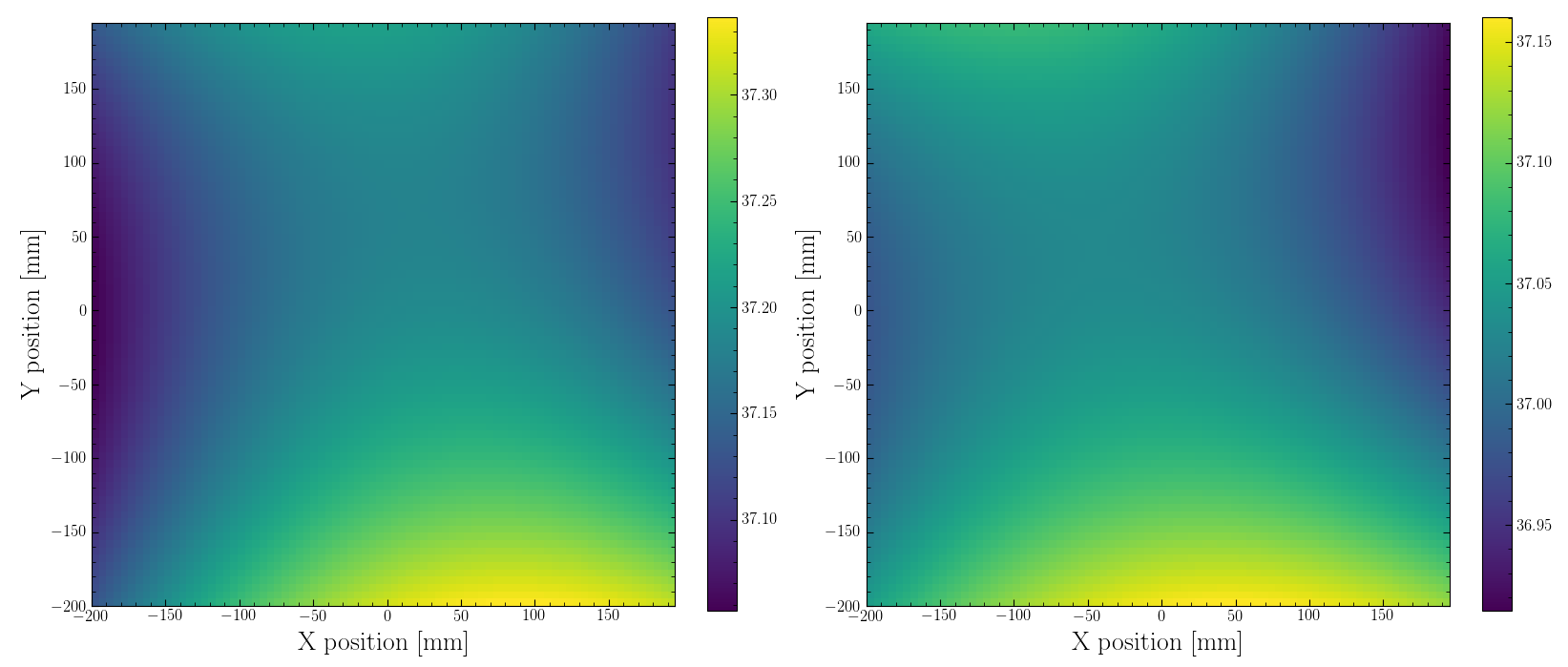}
    \caption{The field plate z-height as a function of position for $\left( \emph{left}\right)$ Morta and $\left( \emph{right}\right)$ Nona. All measurements were made using a DTI gauge attached to each robot, for a consistent set of coordinates across the plate. As a result these measurements are relative to the robot gantries.}
    \label{zmaps}
\end{figure}

A rotational offset was measured at (0,0) between the two theta angles, $\pm90^{\circ}$, so that they can be shifted into the same frame. This method was verified by measuring a series of points at both orientations that the robot could reach. The result was consistent with the offset measured at the origin, allowing us to use this as a global offset between the two rotational frames.

The base map was set to be the same as the zero-point used for the flexure metrology, at ZD $ 40^{\circ}$ with a rotation angle of $0^{\circ}$.  The final base maps for the positioner are shown in figure \ref{zmaps}.

There has been a significant change in the Z-map prior to the positioner being shipped to the observatory. This is expected due to the change in distribution of the field plate shims, and also the affect of having the reinforcement strut attached to the centre of the retractor support disc. We see that the difference between Nona and Morta, in figure \ref{zmaps}, is consistent for both plates.

\subsection{Park heights}

The method for measuring the park heights has remained unchanged from Hughes et al.\cite{Hughes}, as this is still the most effective procedure available. Using the dial gauge for this process is not possible due to the restrictions on the robots movement whilst it is mounted. As the fibres must be moved before the height of the porch can be found, using the grippers directly to measure the values requires the least physical intervention. 

\begin{figure}[ht!]
    \centering
    \includegraphics[width=12cm]{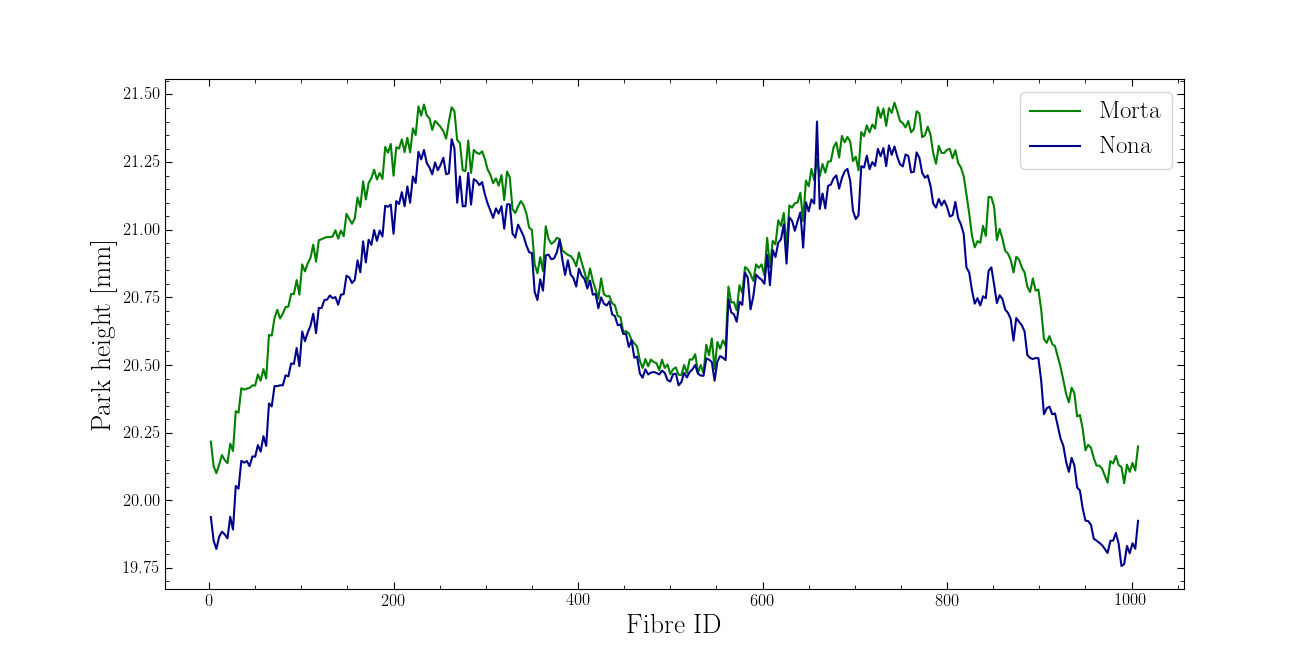}
    \caption{The complete set of park height measurements for plate A, for both Nona and Morta, for all tier 3 fibres.}
    \label{full_parks}
\end{figure}

At ZD $40^{\circ}$, measurements for the complete set of park heights were made for every fibre, across both plates and using both robots. This is shown in figure \ref{full_parks}. The distribution of the park heights remains relatively unchanged compared the values in Oxford, showing the same trend discussed in Hughes et al.\cite{Hughes}. The retractors position has a small tolerance, therefore we do not expect each fibres park height to have shifted significantly, despite the inspection and re-installation of the entire sub-system.

\subsection{Elevation and rotation mapping}

\begin{figure}[ht!]
    \centering
    \includegraphics[width=0.8\linewidth]{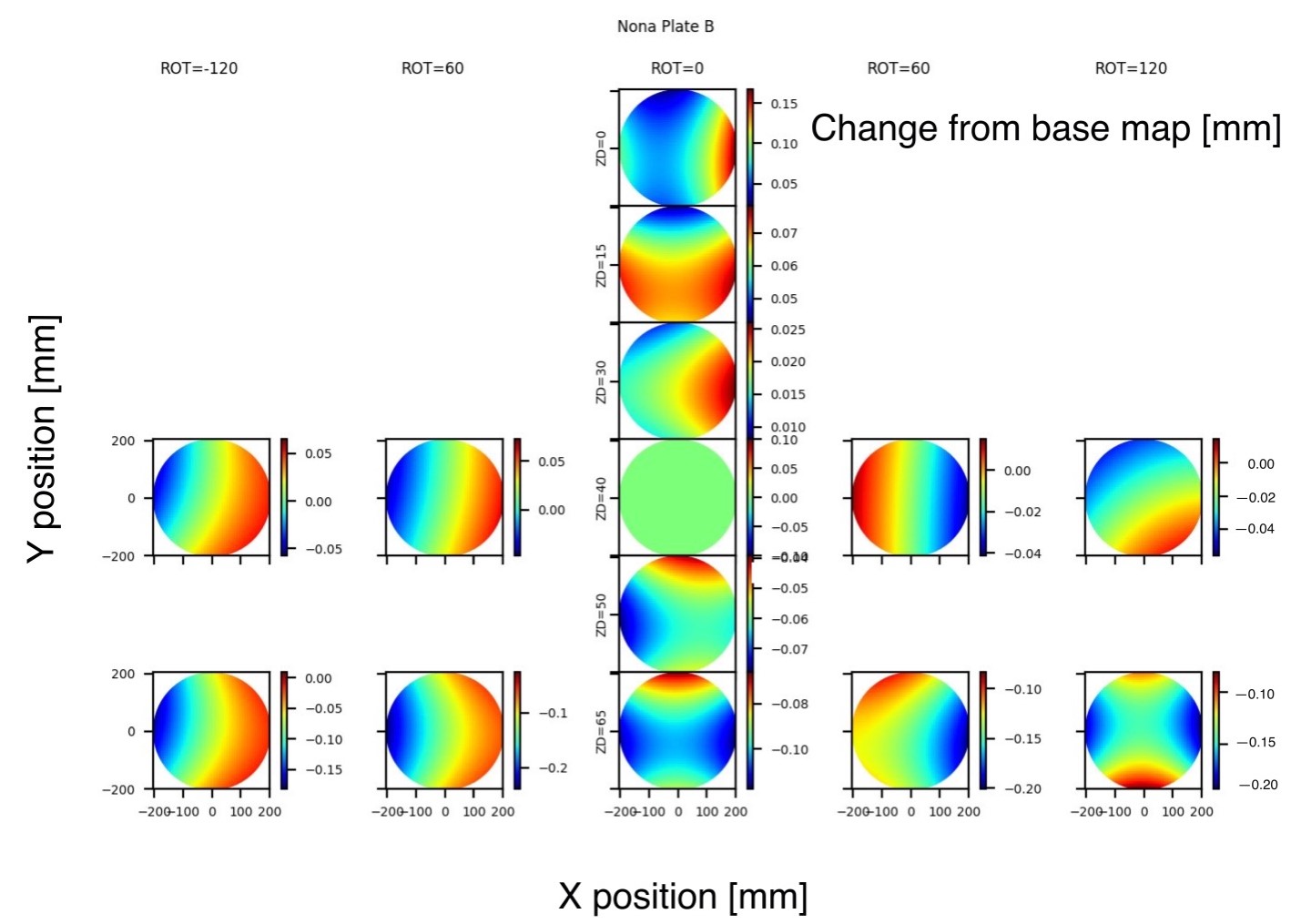}
    \caption{The plate B Nona z-height maps, for a set of elevation and rotation angles which span the range of the orientations that will be used on-sky. These plots show the change in height of the field plate relative to the base map measured at ZD $40^{\circ}$. The individual plots are the difference between the plate maps at rotation angles $+120^{\circ}$, $+60^{\circ}$, $-60^{\circ}$ and ZD angles $0^{\circ}$, $15^{\circ}$, $30^{\circ}$, $40^{\circ}$, $50^{\circ}$, and $65^{\circ}$. The colour represents the change in $z$-height of the plate from the map at ZD $40^{\circ}$.}
    \label{elevation_plates}
\end{figure}

To ensure that the fibres are being picked up and placed accurately during observations, we must understand how the field plates and park heights change as a function of elevation and rotation. By using a dial gauge to map the plates, this procedure can be completed at a range of positions with relative ease by driving the robots remotely. The set of plate maps covering the full range of allowed rotations and elevations is shown in figure \ref{elevation_plates}.

Due to the reduced physical access to the positioner as it changes position, both on the testing rig and on the telescope, we devised a method of extrapolating the z-maps to the park positions in order to predict how they will evolve. 

We begin by defining a function which describes our base plate map $f\left(x,y\right)$. For the complete set of park measurements we have another function called $M\left(X,Y\right)$. The plate map is extrapolated to estimate the plate map at the same position the park heights were measured, $f(X,Y)$. For every park position, we find the difference $\Delta M$, between our extrapolated base map and the park heights, where
\begin{equation}
    \Delta M(X,Y) = f\left(X,Y\right) - M\left(X,Y\right).
\end{equation}
We then apply $\Delta M$ to the new plate map measured for the change in position, which has also been extrapolated to the park positions. This gives a new set of predicted park heights $F\left(X, Y\right)$ that we can then compare to a subset of park heights measured using the gripper jaws at the new position, where
\begin{equation}
    F\left(X, Y\right) = f_{new}\left(X, Y\right) + \Delta M\left(X, Y\right),
\end{equation}
and $f_{new}\left(X, Y\right)$ is the new extrapolated plate map.

For a comparison, direct measurements of the guide retractor porches were made for the same set of positions as the new maps. With plate A, this does not require moving any fibres. However, for plate B the guide retractors are populated, so the tier 3 fibres need to be moved out onto the plate to clear the porches for measurements. The guide fibres were chosen due to their even spacing along the perimeter, and the reduced number of fibres that need to be moved. 

\begin{figure}[ht!]
    \centering
    \includegraphics[width=0.75\linewidth]{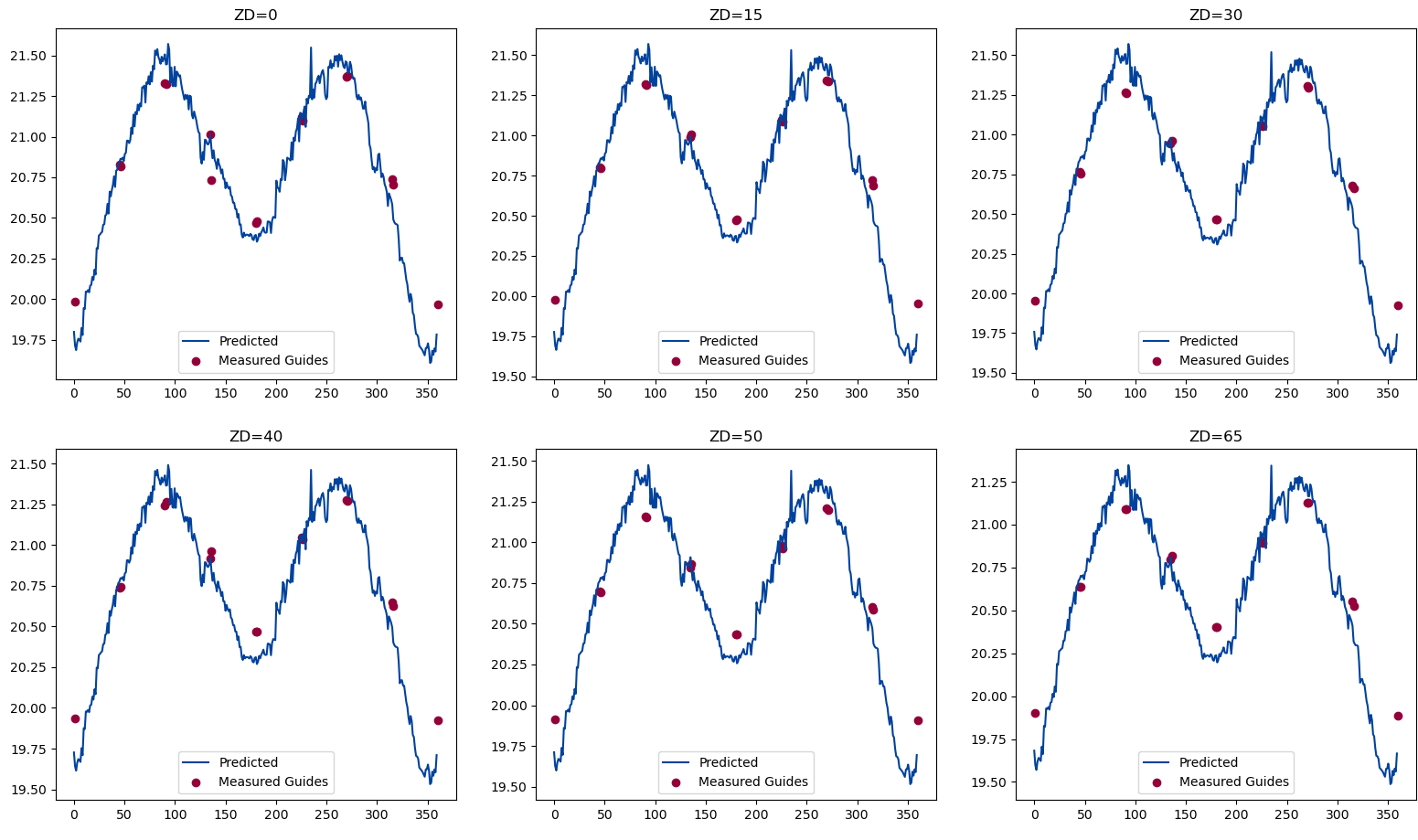}
    \caption{Plots of predicted park heights for Nona on plate A, calculated using the plate maps measured at each ZD angle and the offset between the park heights and plate map at ZD $40^{\circ}$. These predictions were calculated for ZD angles $0^{\circ}$, $15^{\circ}$, $30^{\circ}$, $40^{\circ}$, $50^{\circ}$, and $65^{\circ}$.The red points represent the tier 3 guide retractor heights, which were measured directly using the robot gripper jaws for comparison.}
    \label{elevation}
\end{figure}


\begin{figure}[ht!]
    \centering
    \includegraphics[width=0.75\linewidth]{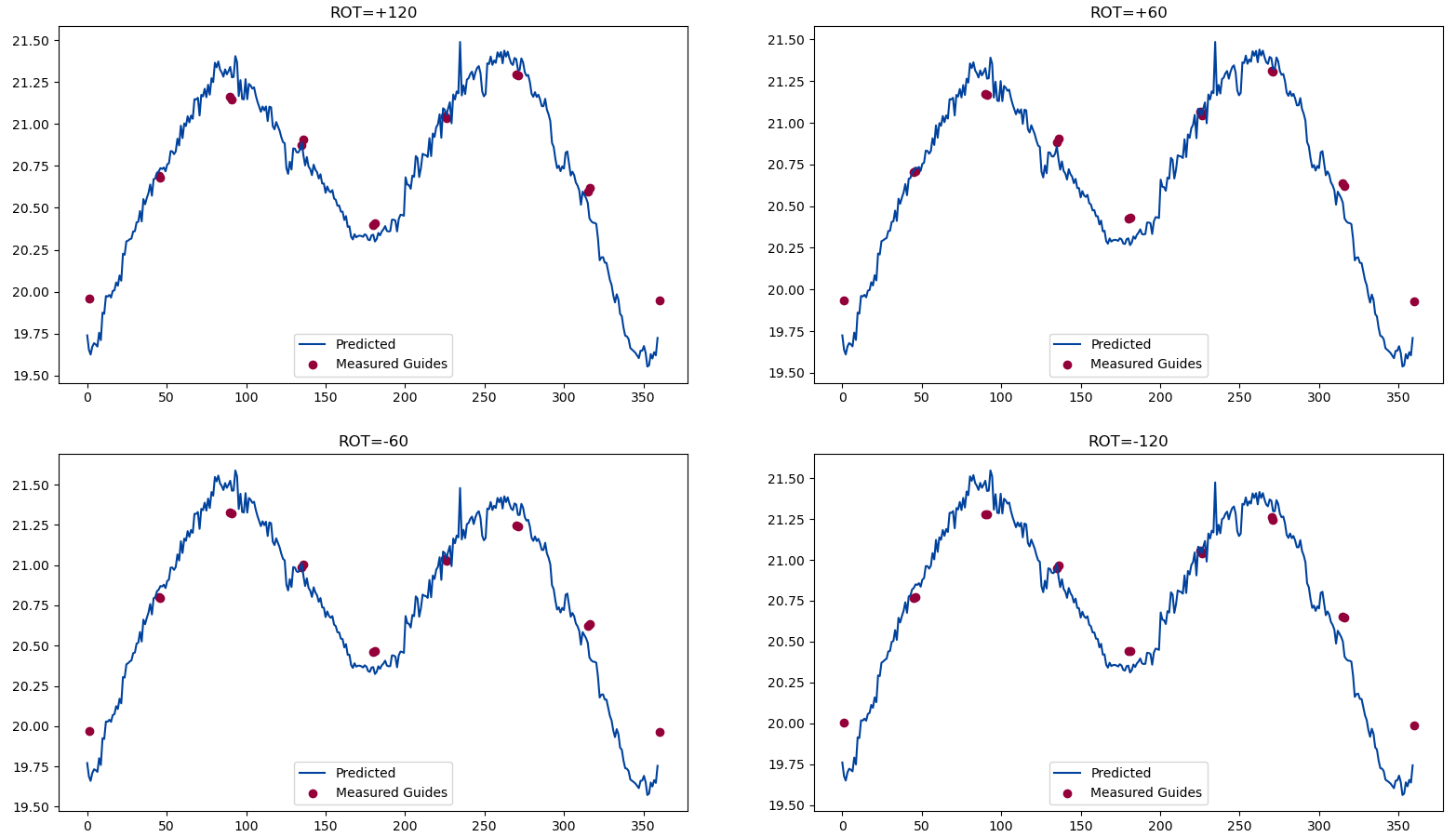}
    \caption{Plots of predicted park heights for Nona on plate A, calculated using the plate maps measured at each rotation angle and the offset between the park heights and plate map rotation angle $0^{\circ}$. These predictions were calculated for rotation angles $+120^{\circ}$, $+60^{\circ}$, $-60^{\circ}$, and $-120^{\circ}$. The red points represent the tier 3 guide retractor heights, which were measured directly using the robot gripper jaws for comparison.}
    \label{rotation}
\end{figure}
\newpage
The change in rotation and ZD are evaluated separately. By understanding how these two movements evolve independently, we can apply their predictions as transformations to the plate maps and park heights. Once WEAVE is on-sky, comparisons can be made to our current models and eventually will be implemented within the positioner's software. The predictions for each elevation angle are shown in figure \ref{elevation}.

For the rotational movement, we chose to set the positions to $+120^{\circ}$, $+60^{\circ}$, $-60^{\circ}$, and $-120^{\circ}$. Figure \ref{rotation} shows our predictions for the park heights on plate A, using the appropriate Z-map for each rotation, and compares it to the set of guide retractor heights found with the robot gripper jaws directly.



For both the elevation and rotation predictions, there remains an uncertainty on the scale of $\sim 40 \mu$m. This uncertainty is estimated from the difference between the measured guide retractor heights and the predictions at each rotation and ZD angle. The residuals we observe appear to be periodic, and we are in the process of generating a routine that will reduce their magnitude to less than $20$ $\mu$m. During observations, the positioner's software will select the closest map to the position of the telescope when each field is configured. The predicted heights for each fibre can then be read directly from the corresponding map file. This will require additional maps to be taken once the positioner has been mounted onto the telescope, as well as validating this approach when combining both rotations and ZD predictions.    

\section{Calibration procedures}

With the positioner's updated metrology, the fibre movement processes must be re-calibrated. This includes calculating the new focus height and rotation centres for all three cameras on each robot.


\subsection{Robot calibration}

We begin by setting the zero point of the encoders for both robots. This is done by either finding the centre spot, or by placing a fibre by hand over the central spot and moving the robot so that the fibres position can be found using the positioner's centroiding software. With the robots in position the encoder values are re-zeroed.

After being reset, the $Z$ height of the robot is adjusted so that the fibre is in focus with the gripper camera. We then calculate the axis of rotation for all three cameras on each robot. This procedure has been refined, after finding that our original method did not evaluate this number to a high enough accuracy. 

A fibre is placed on the rotation centre of the gripper camera by hand as accurately as possible, and the gripper camera measures its position. The view is switched to the upper camera, and the fibre is located in its field of view. This value will be different to that of the gripper camera. The upper camera is then moved to a sequence of four points, which form the corners of a square around the position of the fibre. At each point the camera is kept still whilst the fibres position is re-measured. The difference in the fibres position between all corners is calculated. This is used to set the plate scale of the camera, until the measured position at all four points agree with one another. The rotation centre of the upper camera is then set in pixels, so that it agrees with the position of the fibre as measured by the gripper camera.

After setting the camera parameters, a base-grid of the field plate is measured. Both field plates have a series of points etched into their surface, called the grid, which can then be imaged using the camera's attached to each robot. The position of the grid points are measured using a centroiding software, which are then used to remove the offset, rotation, scaling and non-perpendicularity of the axes when placing fibres on the field plate. Further details on the grid measurements and software can be found in Schallig E. (2019)\cite{schallig_2019}.

The process of measuring the fibre's position with each camera and the four-points is repeated, along with another base-grid from the robots park position, to ensure that the measurements are self-consistent. For this entire process, the robot should be kept at the a constant $\theta$ angle, the same as its park position, since it was found that changing the robots orientation gave inconsistent values. This is likely due to the relative tilt of the robot axes.

It is crucial to ensure that the robots optical system agrees on the fibre's position for both foci. For the gripper focus this should be to within $\pm1$ $\mu$m. However, for the upper focus this can be within $\pm5$ $\mu$m. The upper focus can have a lower tolerance as it is mostly used when searching for misplaced fibres. As long as the fibre is within the field of view, then this feature will be functional. Almost all fibre movements and many of the calibration procedures are made using measurements that are relative to the gripper focus, therefore the tolerances must be as small as possible.

From this point, we can begin running scans of the field plate grids across the same set of orientations as the field plate maps. The first grid tests were ran for the ZD $40^{\circ}$ zero point position for both plates. As before with the field plate maps, we change the rotation and elevation angles independently, so that a complete set of grids can be measured for each position. These will be used as the default in software, which will then be replaced when the grids are measured during the field configuration process on-sky.

\subsection{Fibre placement offsets}

The fibre placement offsets can be divided into four main components; The grasp offsets, the plate release offsets, the park release offsets, and the unpark release offsets. These describe all the stages in a fibres movement cycle.

The grasp offset is the change in the position of the fibre once the gripper jaws have closed around the button vane. Specifically, this is the change in position of the fibres centroid, relative to its button vane. Factors that can affect this include the shape of the vane due to the manufacturing process, the location of the prism relative to the front of the button, and the height at which the fibre is being picked up from.

\begin{figure}[ht!]
    \centering
    \includegraphics[width=0.8\linewidth]{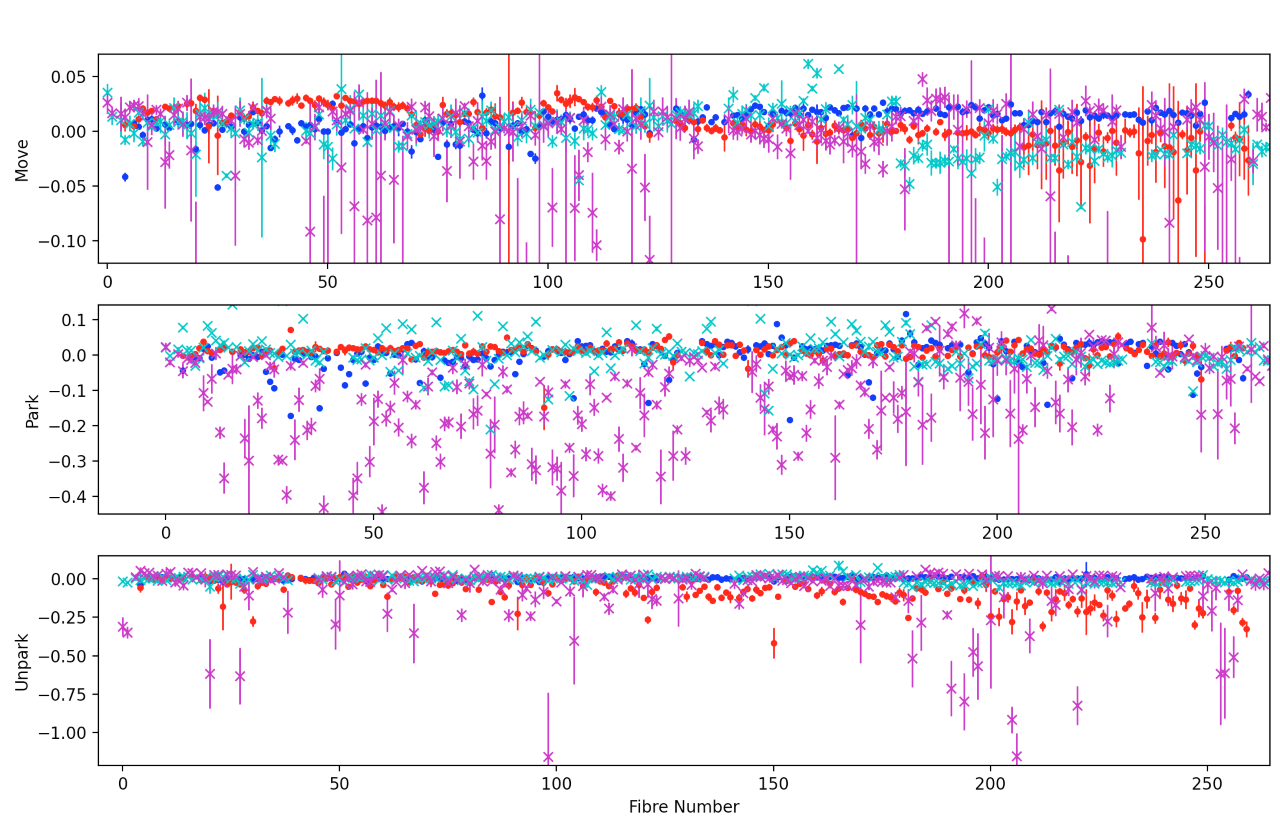}
    \caption{The Move $\left(top \right)$, Park $\left(middle \right)$, and Unpark $\left(bottom \right)$ offsets for 250 fibres on plate A. Here, the cyan markers indicate the offset in the $x$ direction, and magenta markers represent the $y$ direction offset. These measurements have been made using a gripper force of 30 N. The blue and red points represent the $x$ and $y$ direction offsets after the gripper force has been increased to 50 N. }
    \label{offsets}
\end{figure}

The park release offset is found by measuring the shift in position of the fibre when it is released onto its porch, referred to as its park position. This is then used to adjust the positioning of the robot over the porch when it goes to place the fibre back to its resting position. The unpark release offset follows a similar process, except it is the movement of the fibre that has been carried from its park position and placed onto the field plate.

The plate release offset, referred to as \emph{Move} in figure \ref{offsets}, is the change in position of the fibre when it has been carried from one position on the field plate to another. The reason for these different offsets is that the relative tilt angle of the plate to the robots, and each retractor porch to the robot is different. Whilst these angles are relatively small, they have a significant impact on the fibres coordinate shift.

In previous offset measurements, each fibre completed 50 moves on the field plate, which were randomly distributed within it's patrol range\cite{Hughes}. We learnt that when the positioner is in its zenith position, the release offsets do not change as a function of position on the field plate. This must also be verified for a range of elevations. 

We can see the distribution of the mean offsets for each fibre at ZD $40^{\circ}$ in figure \ref{offsets}, which have been measured for 250 fibres. These measurements are relative to the grasp offset calculated for each fibre. This procedure is still ongoing, as the initial measurements, shown by the magenta ($y$ direction offset) and cyan markers ($x$ direction offset), had large random errors. To reduce the magnitude of the offsets, the gripper force was increased from 30 N to 50 N. The offsets measured after this increase are represented by the red ($y$ direction offset) and blue ($x$ direction offset) markers in the figure \ref{offsets}. From this plot, we can see that increasing the gripper force has significantly reduced the magnitude of the offsets random errors. However, we still observe some unusual behaviour in the offsets $y$ direction which requires closer inspection. The reduction of these errors also allows for more anomalous points to be identified and investigated further.
\section{Conclusion}

The new top-end of WEAVE has successfully been installed, and all its components have now been aligned. The observatory is in the process of conducting operational tests, and the commissioning period is due to begin by August 2022.  

\begin{figure}[ht!]
    \centering
    \includegraphics[width=12cm]{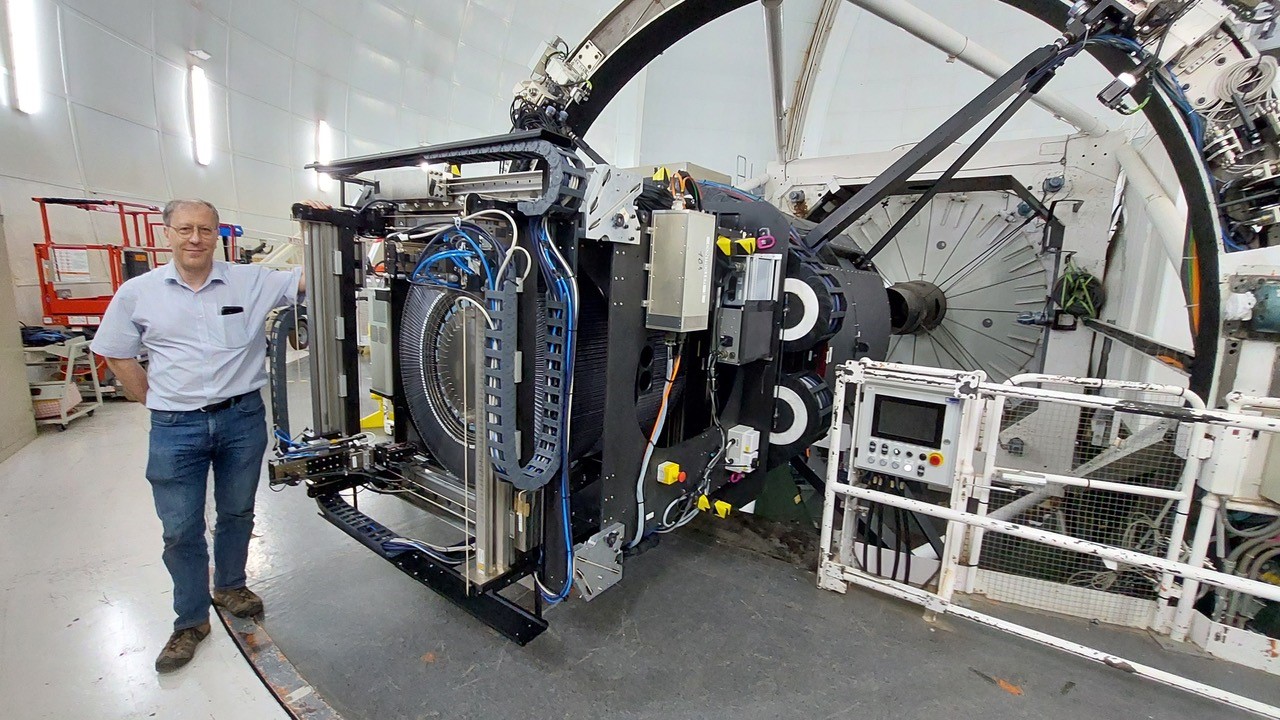}
    \caption{The new top-end of the WHT, positioned at AP3}
    \label{gavin_pic}
\end{figure}

The positioner exhibited significant flexure behaviour with respect to the WRS interface plate, which was outside the tolerance of defocus across each fibres button. Several sets of measurements were taken with both a laser tracker and multiple DTI's to trace the source of the flexure, which was identified to be the tumbler beam and the top plates attached to it. After installing a clamp, aluminium stiffening structures, and two pistons that connect to the RSD, the magnitude of the flexure behaviour was reduced to a maximum displacement of 200 $\mu$m. In addition, both plates now follow the same function of displacement with elevation.

Despite these measures, we still observe a movement of the positioner with respect to the WRS interface, as the rotation angle is changed. When calculating the affect of the defocus on the PSF, the worst case scenario was found to be $1.2"$ at $+150$ $\mu$m defocus.

After the maintenance and flexure work was completed, all 336 retractors installed in the positioner were extracted, inspected, and repaired if necessary. This process took approximately 5 weeks to complete, and resulted in many disabled fibres, including guide fibres, being brought back online. There have been some retractor failures since this work, however this is to be expected. The setup of each retractor can easily be disturbed, and the act of removing and re-installing them is enough to do so.

After re-assembling the positioner, the metrology of the field plates was re-measured using a DTI attached to each robot. Maps were taken for a series of elevation and rotation angles to cover the full range of orientations that will be used during observations. Predictive models were generated for the park heights of every fibre across the set of orientations for the positioner, using the complete set of park measurements and the plate map at ZD $40^{\circ}$ as its basis. These predictions will be adjusted when on-sky, and currently serve as a starting point when performing the other calibration procedures. They have residuals of order $\pm40$ $\mu$m, prior to further analysis.

The robots rotation centre was carefully measured for all three camera's so that they are in agreement to within a few micrometers. For the gripper camera, this measurement has a much smaller tolerance, as all the fibre movements are made with respect to its measurement of the fibre's position. The new procedure for calculating this axis has proven to be accurate enough that it may be used during configuration testing.

Finally, the Move, Park, and Unpark offsets of 250 fibres have been measured on plate A, and this process is still ongoing. It was found that increasing the robots gripper force from its default value of 30 N to 50 N, caused a substantial reduction in the random errors observed. Some unusual behaviour is still seen in the offsets $y$ direction, which will be investigated further after the remaining offset measurements are made. At this stage, the complete set of offsets have been measured for plate A, and are now being conducted for plate B.

\acknowledgments 
 
This work was supported by the Science and Technology Facilities Council and St. Cross College, Oxford. The author would like to give special acknowledgement to SPIE for their support in funding their attendance to this conference.

Funding for the WEAVE facility has been provided by UKRI STFC, the University of Oxford, NOVA, NWO, Instituto de Astrofísica de Canarias (IAC), the Isaac Newton Group partners (STFC, NWO, and Spain, led by the IAC), INAF, CNRS-INSU, the Observatoire de Paris, Région Île-de-France, CONCYT through INAOE, Konkoly Observatory of the Hungarian Academy of Sciences, Max-Planck-Institut für Astronomie (MPIA Heidelberg), Lund University, the Leibniz Institute for Astrophysics Potsdam (AIP), the Swedish Research Council, the European Commission, and the University of Pennsylvania.  The WEAVE Survey Consortium consists of the ING, its three partners, represented by UKRI STFC, NWO, and the IAC, NOVA, INAF, GEPI, INAOE, and individual WEAVE Participants. The WEAVE website can be found at \url{https://ingconfluence.ing.iac.es/confluence//display/WEAV/The+WEAVE+Project} and the full list of granting agencies and grants supporting WEAVE can be found at \url{https://ingconfluence.ing.iac.es/confluence/display/WEAV/WEAVE+Acknowledgements}.

\end{document}